\documentclass{article}
\usepackage[utf8]{inputenc}
\usepackage{amsmath}
\usepackage{amssymb}
\usepackage[title]{appendix}
\usepackage{authblk}
\usepackage[]{caption} 
\usepackage{cite}
\usepackage{comment}
\usepackage{enumitem}
\usepackage{etoolbox}
\usepackage{geometry}
\usepackage{graphicx}
\usepackage{hyperref}
\usepackage{longtable}
\usepackage{mathtools}
\usepackage{multicol}
\usepackage{nomencl}
\usepackage{siunitx}
\usepackage{subcaption}
\usepackage{tabularx}
\usepackage{url}

\newcommand{\B}{\vec{B}}
\newcommand{\Ba}{\B_a} 
\newcommand{\Be}{\B_e} 
\newcommand{\Bt}{\B_t} 

\newcommand{\bpf}[1]{\mathrm{bpf} (#1)}

\newcommand{\epsnot}{\epsilon_0}
\newcommand{\ddt}[1]{\frac{d}{dt}#1}
\newcommand{\mat}[1]{\boldsymbol{#1}}
\newcommand{\munot}{\mu_0}

\graphicspath{{figs/}}

\title{Derivation and Extensions of the Tolles-Lawson Model for Aeromagnetic Compensation}

\author[1*]{Albert R. Gnadt}
\author[2*]{Allan B. Wollaber}
\author[3*]{Aaron P. Nielsen}

\affil[1]{Massachusetts Institute of Technology, Cambridge, MA, USA}
\affil[2]{MIT Lincoln Laboratory, Lexington, MA, USA}
\affil[3]{Air Force Institute of Technology, Wright-Patterson AFB, OH, USA}
\affil[*]{all authors contributed equally}
\date{December 15, 2022}

\begin{document}

\maketitle

\begin{abstract}
\noindent This note is intended to serve as a straightforward reference that summarizes and expands on the linear aeromagnetic compensation model first introduced by Tolles and Lawson in 1950. The Tolles-Lawson model provides a simple, physical representation of an aircraft's magnetic field, composed of permanent, induced, and eddy current terms. It utilizes an approximation (a Taylor expansion) to enable fitting coefficients with a general linear model. Here, the Tolles-Lawson model is derived, paying stricter attention to where assumptions are made, the model calibration procedure is described, and some additional comments on a second-order correction and a means of constructing the vector aircraft field are provided.
\end{abstract}

\section{Introduction and Derivation}

\noindent Tolles and Lawson first reported their linear aeromagnetic compensation model for scalar magnetometers in 1950 \cite{Tolles1950}, though it was developed earlier during World War II. Tolles was later issued patents for the hardware involved in airborne magnetometer compensation and the primary initial use case was magnetic anomaly detection (MAD) \cite{Tolles1954,Tolles1955}. Leliak later proposed performing sinusoidal maneuvers during a calibration flight to increase observability of the terms in the Tolles-Lawson model \cite{Leliak1961}. This has been the state-of-the-art for decades, though numerous improvements to this method have been proposed over the years \cite{Leach1979,Gu2013,Webb2014,Wu2018a}. The basic idea of the Tolles-Lawson model \cite{Bickel1979} is to use magnetic measurements from a vector magnetometer to calibrate a scalar magnetometer, the latter of which is used for navigation. This model provides a means for removing a corrupting aircraft magnetic field from a scalar total magnetic field measurement, yielding the Earth magnetic field used for navigation.
An airborne vector magnetometer measures the vector sum of two primary magnetic fields,

\begin{equation}
    \Bt = \Be + \Ba \, ,
\end{equation}

\noindent where $\Bt$ is the total field, $\Be$ is the Earth (external) field, and $\Ba$ is the unknown aircraft (interference) field. Note that here Earth field refers to all components – core field, anomaly field , and temporal variations (i.e., space weather). A vector magnetometer measures $\Bt$, but for navigation the unknown, desired signal is $|\Be|$, the magnitude of $\Be$. A scalar magnetometer measures $|\Bt|$, the magnitude of $\Bt$. These terms can be related as follows:

\setlength{\jot}{2ex} 
\begin{gather}
    |\Be|^2 = \Be \cdot \Be = (\Bt - \Ba) \cdot (\Bt - \Ba) \, , \\
    |\Be|^2 = \Bt \cdot \Bt - 2 \Ba \cdot \Bt + \Ba \cdot \Ba \, , \\
    |\Be| = \sqrt{|\Bt|^2 - 2 \Ba \cdot \Bt + |\Ba|^2} \, , \\
\label{eq:full_B_t}
    |\Be| = |\Bt| \sqrt{1 - 2 \frac{\Ba \cdot \Bt}{|\Bt|^2} + \frac{|\Ba|^2}{|\Bt|^2}} \, .
\end{gather}

\noindent The Tolles-Lawson model aims to create a physical representation of $\Ba$, which appears in a nonlinear form in \eqref{eq:full_B_t}. In order to create a linear model for the aircraft field, it is possible to leverage the fact that ${|\Ba|}/{|\Bt|}$ is  $O(\epsilon)$, i.e., the aircraft field is small compared to the total field. Using the identity $\Ba \cdot \Bt = |\Ba||\Bt|\cos\theta$  and defining $\epsilon= |\Ba|/|\Bt|$, \eqref{eq:full_B_t} becomes

\begin{gather}
    |\Be| = |\Bt| \sqrt{1 - 2\cos\theta  \frac{|\Ba|}{|\Bt|} + \frac{|\Ba|^2}{|\Bt|^2}} \, , \\
    |\Be| = |\Bt| \sqrt{1 - 2\cos\theta  \epsilon + \epsilon^2} \, ,
\intertext{which can be linearized using a Taylor expansion,}
\label{eq:series}
    \sqrt{1 - 2 \cos \theta \epsilon + \epsilon^2} = 1 - \frac{2\cos \theta}{2}\epsilon
     + (1-\cos^2\theta)\frac{\epsilon^2}{2}
     + O(\epsilon^3) \text{ as } \epsilon \rightarrow 0 \, , \\
    |\Be| = |\Bt| \left(1 - \frac{\Ba \cdot \Bt}{|\Bt|^2} + O\left(\frac{|\Ba|^2}{|\Bt|^2}\right) \right) \, ,
\intertext{which, discarding terms of $O(\epsilon^2)$, gives the linear relationship}
    \label{eq:lin_B_t}
    |\Be| \approx |\Bt| - \frac{\Ba \cdot \Bt}{|\Bt|}.
\end{gather}

\noindent The individual components of the vector magnetometer are used to compute the total field direction cosines,

\begin{equation} \label{eq:dir_cos}
     \hat{B}_t = \frac{\Bt}{|\Bt|}.
\end{equation}

\noindent Using this definition, \eqref{eq:lin_B_t} becomes

\begin{equation} \label{eq:tl_1}
    |\Be| \approx |\Bt| - \Ba \cdot \hat{B}_t \, ,
\end{equation}

\noindent where $|\Be|$ is the magnitude of the Earth field (desired signal for navigation) and $|\Bt|$ is the (measured) total field. Note that $\Ba \cdot \hat{B}_t$ is a corruption term, i.e. the effect of the aircraft field projected onto the total field.

Up to this point, no physics knowledge has been incorporated. The derivation comes solely from manipulating vectors and making an assumption about the magnitude of those vectors. In order to get to the final model, Tolles and Lawson \cite{Tolles1950} assumed that the aircraft field is comprised of permanent, induced, and eddy current magnetic moments,

\begin{equation}
    \Ba = \B_\mathrm{perm} + \B_\mathrm{ind} + \B_\mathrm{eddy} \, ,
\end{equation}

\noindent These quantities can be argued from physical principles starting with the quasi-static model of electrodynamics developed by Darwin \cite{Larsson2007}. A quasi-static model is required because the wavelength in free space for the frequencies of interest ($<10$ Hz) is greater than $f/c = 30,000$ km, where $f$ is the frequency and $c$ is the speed of light, and this wavelength is much larger than an aircraft or other vehicle. In this model, Ampere's law takes a modified form where the displacement current is replaced with a term due only to free charges, neglecting a Faraday-like term,

\begin{equation} \label{eqn:ampere-darwin}
    \nabla\times\vec{H} = \vec{J} - \epsnot\ddt\vec{E_C} \, ,
\end{equation}

\noindent where $\vec{H}$ is the magnetic induction, $\vec{J}$ is the current density, $\epsnot$ is the permittivity of free space and $\vec{E_C}$ is the electric field generated by Coulomb charges as defined in \cite{Larsson2007}. Combining the displacement current $\epsnot\ddt\vec{E_C}$ with Ohm's law $\vec{J}=\mat{\sigma}\vec{E}$ to get

\begin{equation} \label{eq:Ampere_modified}
    \nabla \times \vec{H}_\text{total} = \vec{J}_\text{perm} - \epsnot\mat{\sigma}^{-1}\ddt\vec{J}_\text{eddy} \, ,
\end{equation}

\noindent in which $\vec{J}_\text{perm}$ represents any unchanging currents that could be represented as a permanent moment and $- \epsnot\mat{\sigma}^{-1}\ddt\vec{J}$ represents the eddy currents. In the case of the Tolles-Lawson model, $\vec{J}$ is indistinguishable from a permanent magnetization, hence in terms of time-varying fields only, \eqref{eq:Ampere_modified} becomes

\begin{equation} \label{eq:Ampere_modified_time_varying}
    \nabla \times \vec{H}_\text{eddy} = - \epsnot\mat{\sigma}^{-1}\ddt\vec{J}_\text{eddy} \, .
\end{equation}

\noindent Now consider that a magnetization density $\vec{M}$ may be expressed as a bound current $\vec{J}_b$ as in Griffiths Section 6.2 \cite{Griffiths1999,Herczynski2013} via

\begin{equation}
    \vec{J}_b= \nabla\times \vec{M} \, .
    \label{eqn:bound_current_mag}
\end{equation}

\noindent This relationship may be applied since, in the case of an aircraft, the currents are confined to conducting materials and cannot flow arbitrarily. Solving to get the eddy current generated field $\vec{H}_\text{eddy}$ as a time varying magnetization

\begin{equation}
    \vec{H}_\text{eddy} = - \epsnot\mat{\sigma}^{-1}\ddt\vec{M}_\text{eddy} \, .
\end{equation}

\noindent It is further assumed that these eddy current moments are proportional to the external field which creates them $\vec{M}_\text{eddy} \propto \vec{H}_\text{external}$ via some unknown relationship to the aircraft construction. The magnetic $\B$-field is related to the $\vec{H}$-field through

\begin{equation}
    \B = \munot \left[ \vec{M} + \vec{H} \right] \, .
\end{equation}

\noindent The magnetization model for a ferromagnetic system, as long as the external field remains well below the coercive field strength, is

\begin{equation}
    \vec{M} = \vec{M}_\text{perm} + \mat{\mu} \vec{H}_\text{external} \, ,
\end{equation}

\noindent with permanent magnetization $\vec{M}_\text{perm}$ and induced magnetization related through the external field $\vec{H}_\text{external}$. Finally, the total magnetic field measured by a sensor also includes the eddy current contribution,

\begin{equation} \label{eq:B_to_H_si}
    \B_\text{total} = \munot\vec{H}_\text{external} +  \munot \left[ 
        \vec{M}_\text{perm} + 
        \mat{\mu} \vec{H}_\text{external} - \epsnot\mat{\sigma}^{-1}\ddt\vec{M}_\text{eddy}
        \right] \, .
\end{equation}

\noindent The second portion of \eqref{eq:B_to_H_si},

\begin{equation}
    \Ba = 
    \munot\vec{M}_\text{perm} + 
    \munot\mat{\mu} \vec{H}_\text{external} - \munot\epsnot\mat{\sigma}^{-1}\ddt\vec{M}_\text{eddy} \, ,
\end{equation}

\noindent is the aircraft generated disturbance field that is reduced to the simplified form of

\begin{equation} \label{eq:three_fields}
    \Ba = \boldsymbol{a} + \boldsymbol{b} \Bt + \boldsymbol{c} \dot{\B}_t \, ,
\end{equation}

\noindent where coefficient vector $\boldsymbol{a}$ and coefficient matrices $\boldsymbol{b}$ and $\boldsymbol{c}$ are all unknown, and the $\dot{()}$ notation indicates a time-derivative. Note that, in principle, the $\Bt$ terms should be replaced by $\Be$, as it is the external, Earth magnetic fields that induce $\B_\mathrm{ind}$ and $\B_\mathrm{eddy}$, and not the total, measured field. However, for geomagnetic surveying and during navigation, only $\Bt$ is available. In practice, there is no difficulty in treating \eqref{eq:three_fields} as a definition, which has been the norm for years \cite{Bickel1979, Canciani2021}, and $\Be$ is close enough to $\Bt$ that the physical meaning of this approximation is essentially retained. The permanent magnetic moment terms,

\begin{equation} \label{eq:perm}
    \B_\mathrm{perm} = \boldsymbol{a} = 
    \begin{bmatrix}
        a_1 & a_2 & a_3
    \end{bmatrix}^T \, ,
\end{equation}

\noindent contain 3 unknown coefficients. These permanent magnetic moment terms represent nearly constant, permanent magnetization of various ferromagnetic aircraft components, including both the aircraft itself and items within the aircraft \cite{Bickel1979}. These terms do not change unless the aircraft configuration or contents are modified. The induced magnetic moment terms,

\begin{equation} \label{eq:ind}
    \B_\mathrm{ind} = \boldsymbol{b} \Bt = |\Bt|
    \begin{bmatrix}
        b_{11} & b_{12} & b_{13} \\
        b_{21} & b_{22} & b_{23} \\
        b_{31} & b_{32} & b_{33} \\
    \end{bmatrix}
    \hat{B}_t \, ,
\end{equation}

\noindent contain 9 unknown coefficients. These induced magnetic moment terms represent the Earth field inducing a secondary magnetic field in magnetically susceptible aircraft components. The relative orientation of the aircraft and Earth field determines the magnitude and direction of the induced magnetization. Since much of the aircraft structure is comprised of non-magnetic aluminum alloys, the primary source of induced magnetic fields are the aircraft engines \cite{Reeves2005}. The eddy current terms,

\begin{equation} \label{eq:eddy}
    \B_\mathrm{eddy} = \boldsymbol{c} \dot{\B}_t = |\dot{\B}_t|
    \begin{bmatrix}
        c_{11} & c_{12} & c_{13} \\
        c_{21} & c_{22} & c_{23} \\
        c_{31} & c_{32} & c_{33}  
    \end{bmatrix}
    \hat{\dot{B}}_t \, ,
\end{equation}

\noindent contain 9 unknown coefficients. These eddy current terms represent electrical current loops caused by the time-varying Earth field (relative to the aircraft) interacting with electrically conductive aircraft components. Unlike the permanent and induced fields, eddy currents depend on the time rate of change of Earth's magnetic flux through these components, such as the aircraft skin \cite{Tolles1950, Leliak1961}. Magnetic fields created by eddy currents obey Lenz's law, opposing the magnetic field that created them \cite{Canciani2016a}. This is similar to how current is produced in a coil rotating in a uniform magnetic field \cite{Bickel1979}. The eddy current terms are typically the smallest of the three fields and are sometimes neglected or approximated with diagonal terms. In fact, prior treatments of this term use an unstated approximation that $\dot{\B} \approx |\B|\dot{\hat{B}}$, which works in practice, but is unnecessary \cite{Bickel1979, Canciani2016a}. The form of the aircraft generated moments (corruption term) in \eqref{eq:tl_1} becomes

\begin{equation} \label{eq:tl_1b}
    \Ba \cdot \hat{B}_t = 
    (\boldsymbol{a} + 
    |\Bt| \ \boldsymbol{b} \ \hat{B}_t + 
    |\dot{\B}_t| \ \boldsymbol{c} \ \hat{\dot{B}}_t) \cdot \hat{B}_t \, .
\end{equation}

\noindent There are a total of 21 coefficients in $\boldsymbol{a}$, $\boldsymbol{b}$, and $\boldsymbol{c}$, but due to symmetry in the induced magnetic moment matrix $\boldsymbol{b}$, the repeated off-diagonal terms are removed resulting in 3 fewer coefficients. Thus, there are 18 total unknown coefficients in the standard Tolles-Lawson model,

\small
\begin{equation} \label{eq:tl_2}
    |\Be| \approx |\Bt| - \left( \hat{B}_t^T
    \begin{bmatrix}
        \beta_1 \\
        \beta_2 \\
        \beta_3 \\
    \end{bmatrix} + 
    |\Bt| \ \hat{B}_t^T
    \begin{bmatrix}
        \beta_{4} & \beta_{5} & \beta_{6} \\
        \cdot     & \beta_{7} & \beta_{8} \\
        \cdot     & \cdot     & \beta_{9} \\
    \end{bmatrix}
    \hat{B}_t + 
    |\dot{\B}_t| \ \hat{B}_t^T
    \begin{bmatrix}
        \beta_{10} & \beta_{11} & \beta_{12} \\
        \beta_{13} & \beta_{14} & \beta_{15} \\
        \beta_{16} & \beta_{17} & \beta_{18} \\
    \end{bmatrix}
    \hat{\dot{B}}_t \right) \, ,
\end{equation}
\normalsize

\noindent which has unknowns on both sides of the equation, $|\Be|$ and $\boldsymbol{\beta}$, since only the total field $\Bt$ and $|\Bt|$ can be directly measured during geomagnetic surveying. However, using a ``trick'' \eqref{eq:tl_2} can be modified into a solvable form even without knowing $|\Be|$.

\newpage

\section{Calibrating the Aircraft Field}

\noindent Rewriting \eqref{eq:tl_2}, first a length 18 row vector of direction cosine terms, which are calculated from a vector magnetometer measurement, is created as

\begin{equation} \label{eq:tl_A1}
    \vec{\delta} = 
    \begin{bmatrix}
    \hat{B}_{\ 3\times1} \\
    \mathrm{vec}(|\B| \ \hat{B} \hat{B}^T)_{6\times1} \\
    \mathrm{vec}(|\dot{\B}| \ \hat{B} \hat{\dot{B}}^T)_{9\times1} \\
    \end{bmatrix}^T \, ,
\end{equation}

\noindent where again only 6 of the induced magnetic moment terms are taken from $|\B| \ \hat{B} \hat{B}^T$ due to symmetry. Explicitly, the 18 direction cosine terms are

\begin{equation} \label{eq:tl_explicit}
    \vec{\delta} = 
    \begin{bmatrix}
    \hat{B}_x \\[0.04cm]
    \hat{B}_y \\[0.04cm]
    \hat{B}_z \\[0.04cm]
    |\B| \hat{B}_x \hat{B}_x \\[0.06cm]
    |\B| \hat{B}_x \hat{B}_y \\[0.06cm]
    |\B| \hat{B}_x \hat{B}_z \\[0.06cm]
    |\B| \hat{B}_y \hat{B}_y \\[0.06cm]
    |\B| \hat{B}_y \hat{B}_z \\[0.06cm]
    |\B| \hat{B}_z \hat{B}_z \\[0.06cm]
    |\dot{\B}| \hat{B}_x \hat{\dot{B}}_x \\[0.05cm]
    |\dot{\B}| \hat{B}_x \hat{\dot{B}}_y \\[0.05cm]
    |\dot{\B}| \hat{B}_x \hat{\dot{B}}_z \\[0.05cm]
    |\dot{\B}| \hat{B}_y \hat{\dot{B}}_x \\[0.05cm]
    |\dot{\B}| \hat{B}_y \hat{\dot{B}}_y \\[0.05cm]
    |\dot{\B}| \hat{B}_y \hat{\dot{B}}_z \\[0.05cm]
    |\dot{\B}| \hat{B}_z \hat{\dot{B}}_x \\[0.05cm]
    |\dot{\B}| \hat{B}_z \hat{\dot{B}}_y \\[0.05cm]
    |\dot{\B}| \hat{B}_z \hat{\dot{B}}_z \\[0.05cm]
    \end{bmatrix}^T \, ,
\end{equation}

\noindent where $\hat{B}_x$, $\hat{B}_y$, and $\hat{B}_z$ are the direction cosines. A time series of $\vec{\delta}$ can be composed into an $N \times 18$ matrix

\begin{equation} \label{eq:tl_A2}
    \boldsymbol{A} = \begin{bmatrix} \vec{\delta}_1 \\ \vdots \\ \vec{\delta}_{N} \end{bmatrix} \, ,
\end{equation}

\noindent where each row is one of $N$ time steps. The column vector of Tolles-Lawson coefficients to learn is $\boldsymbol{\beta}$, as taken from \eqref{eq:tl_2}. Rearranging and substituting, 

\begin{equation} \label{eq:tl_3}
    \boldsymbol{B}_{\mathrm{scalar}} - |\vec{\boldsymbol{B}}_e| = \boldsymbol{A} \beta \, ,
\end{equation}

\noindent where $|\vec{\boldsymbol{B}}_e|$ and $\boldsymbol{\beta}$ are both still unknown and $\boldsymbol{B}_{\mathrm{scalar}}$ is scalar magnetometer measurements that represent $|\vec{\boldsymbol{B}}_t|$. The ``trick'' is to use a bandpass filter (bpf) on \eqref{eq:tl_3}, typically a Butterworth infinite impulse response filter is chosen \cite{Han2017},

\begin{equation} \label{eq:tl_4}
    \bpf{\boldsymbol{B}_{\mathrm{scalar}} - |\vec{\boldsymbol{B}}_e|} = \bpf{\boldsymbol{A} \boldsymbol{\beta}} \, .
\end{equation}

\begin{figure}[ht]
    \centering
    \includegraphics[width=0.45\textwidth]{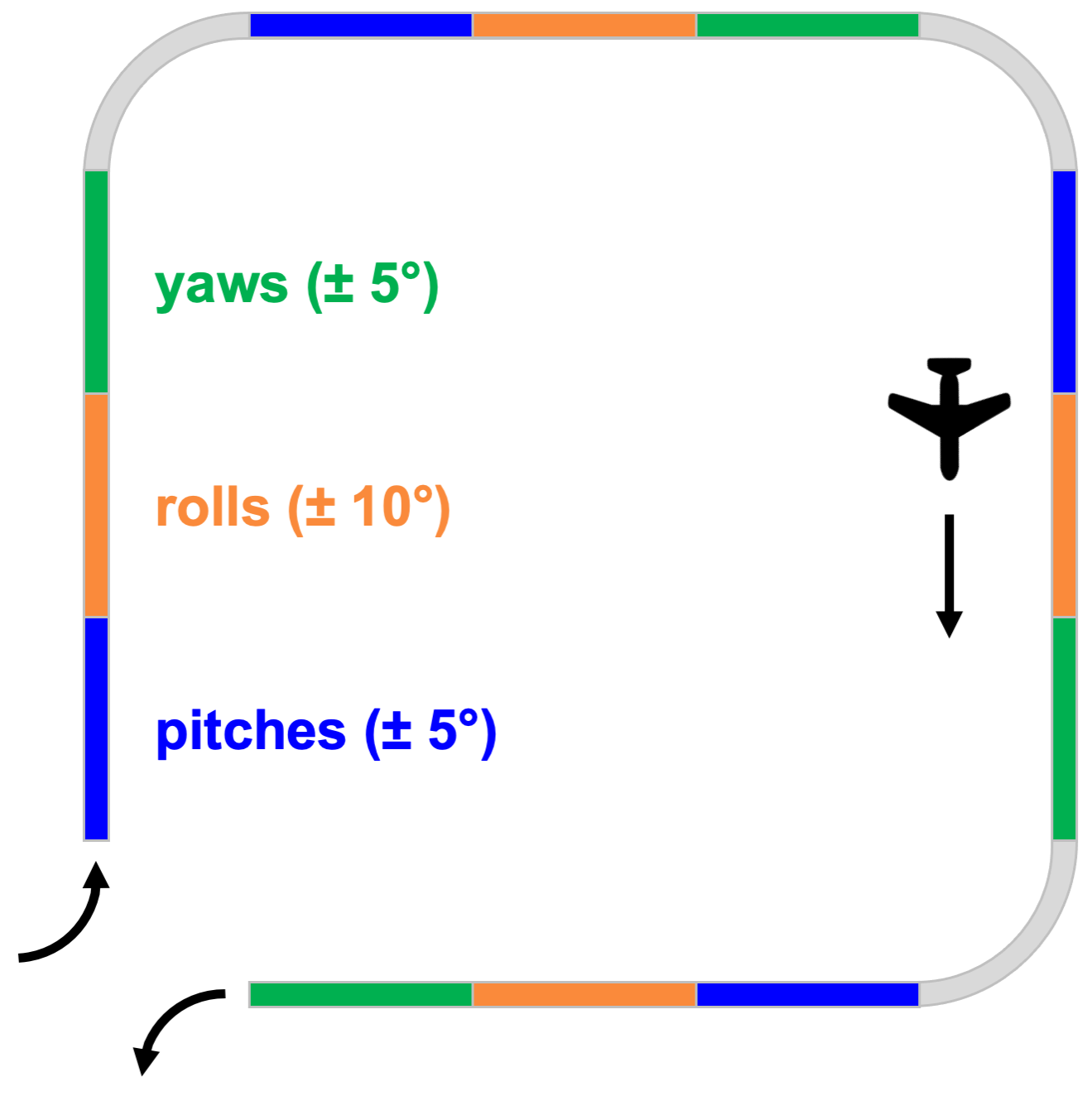} 
    \caption[Typical aeromagnetic calibration flight pattern]{Typical aeromagnetic calibration flight pattern. A box-like flight path with tight turns is flown with pitch, roll, and yaw maneuvers (in any order) performed along each leg at a specific period (usually $1-10$ seconds) \cite{Noriega2011}.}
    \label{fig:cal_path}
\end{figure}

\begin{figure}[ht]
    \centering
    \includegraphics[width=0.55\textwidth]{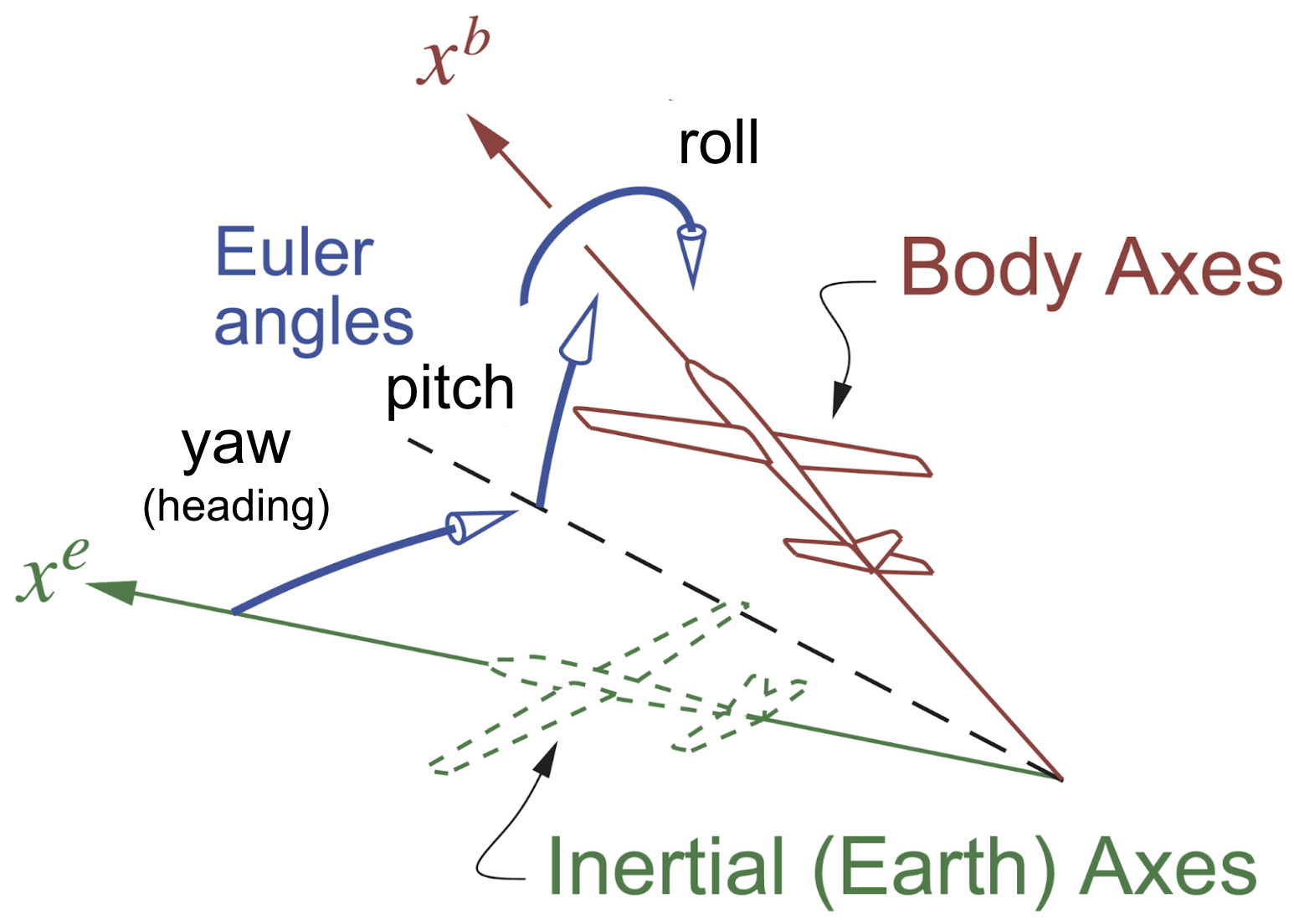} 
    \caption[Euler angles]{Euler angles. Modified from \cite{Drela2013}.}
    \label{fig:euler angles}
\end{figure}

\noindent The passband frequency range for the bandpass filter is carefully selected in order to remove nearly all of the Earth field while keeping much of the aircraft field by setting to the periodicity of the pitch, roll, and yaw maneuvers. In practice, a passband of $0.1 - 0.9$ Hz has been found to perform well, since in this range the frequency content of the aircraft dominates the magnetic signal. The measurements themselves are taken during a specific set of roll, pitch, and yaw aircraft maneuvers during a calibration flight, as shown in Figure~\ref{fig:cal_path}. Roll, pitch, and yaw, as shown in Figure~\ref{fig:euler angles}, are the Euler angles that describe the aircraft orientation in reference to the Earth.

\noindent These maneuvers, are meant to ``inject'' aircraft magnetic field content into the passband frequency range. The calibration flight is performed at a high altitude over a region with a small magnetic gradient to reduce the uncertainty imparted by the Earth field \cite{Bickel1979a,Noriega2011}. When a high-quality map is available for calibration, a bandpass filter is not necessary, and calibration flights should occur at lower altitudes for map-based aeromagnetic calibration and compensation \cite{Canciani2021}. A low-pass filter can still be used to remove high-frequency information that is unrelated to the map. Continuing with $\bpf{|\vec{\boldsymbol{B}}_e|} \approx 0$, \eqref{eq:tl_4} becomes

\begin{equation} \label{eq:tl_5}
    \bpf{\boldsymbol{B}_{\mathrm{scalar}}} = \bpf{\boldsymbol{A}} \boldsymbol{\beta} \, ,
\end{equation}

\noindent where $\boldsymbol{B}_{\mathrm{scalar}}$ is known from scalar magnetometer measurements and $\boldsymbol{A}$ is known from vector magnetometer measurements, as described previously. The Tolles-Lawson coefficients $\boldsymbol{\beta}$ can then be solved for with linear least squares regression,

\begin{equation} \label{eq:tl_llsr}
    \boldsymbol{\beta} = (\boldsymbol{A}_f^T \boldsymbol{A}_f)^{-1} \boldsymbol{A}_f^T \boldsymbol{y} \, ,
\end{equation}

\noindent or ridge regression,

\begin{equation} \label{eq:tl_rr}
    \boldsymbol{\beta} = (\boldsymbol{A}_f^T \boldsymbol{A}_f + \lambda \boldsymbol{I})^{-1} \boldsymbol{A}_f^T \boldsymbol{y} \, ,
\end{equation}

\noindent where $\boldsymbol{y}=\bpf{\boldsymbol{B}_{\mathrm{scalar}}}$, $\boldsymbol{A}_f=\bpf{\boldsymbol{A}}$, and $\lambda$ is a ridge parameter. Ridge regression is useful when $\boldsymbol{A}_f^T \boldsymbol{A}_f$ is poorly conditioned due to correlations among the Tolles-Lawson coefficients \cite{Leach1979}. A ridge parameter of $0.00025 - 0.025$ is appropriate for this type of problem \cite{Nelson2020,Gnadt2022b}. Though the coefficients are determined using bandpass filtered measurements, they can be applied to unfiltered measurements. During a non-calibration flight, compensation is then performed as

\begin{equation} \label{eq:tl_6}
    |\vec{\boldsymbol{B}}_e| = \boldsymbol{B}_{\mathrm{scalar}} - \boldsymbol{A} \boldsymbol{\beta} \, ,
\end{equation}

\noindent where the $\boldsymbol{\beta}$ Tolles-Lawson coefficients represent the average aircraft magnetic field contributions predetermined during a calibration flight and $\boldsymbol{B}_{\mathrm{scalar}}$ and $\boldsymbol{A}$ come from real-time scalar and vector magnetometer measurements, respectively.

\section{A Second-Order Accurate Expansion}

Rather than dropping the second-order term in \eqref{eq:full_B_t} and linearizing \eqref{eq:series} to first order, a second-order solution may be obtained by dropping terms of $O(\epsilon^3)$. To see this, first use the identity

\begin{equation}
    1 - \cos^2\theta = \sin^2\theta = \frac{|\Ba \times \Bt|^2}{|\Ba|^2 |\Bt|^2}
\end{equation}

\noindent in \eqref{eq:series} to obtain

\begin{equation} \label{eq:second_1}
    |\Be| = |\Bt| \left(1 - \frac{\Ba \cdot \Bt}{|\Bt|^2} 
    + \frac{|\Ba \times \Bt|^2}{2 |\Bt|^4}
    + O\left(\frac{|\Ba|^3}{|\Bt|^3}\right) \right) \, .
\end{equation}

\noindent  Using the direction cosine definition in \eqref{eq:dir_cos}, dropping terms of $O(\epsilon^3)$, and rearranging, \eqref{eq:second_1} becomes 

\begin{equation} \label{eq:second_2}
    |\Be| \approx |\Bt| - \Ba \cdot \hat{B}_t + \frac{|\Ba \times \hat{B}_t|^2}{2 |\Bt|} \, ,
\end{equation}

\noindent which requires both $\Ba \cdot \hat{B}_t$ and $\Ba \times \hat{B}_t$. This is an interesting result in that the first and second order corrections are orthogonal to each other. Alternatively, since

\begin{equation} \label{eq:second_3}
    |\Ba \times \hat{B}_t|^2 = |\Ba|^2 - (\Ba \cdot \hat{B}_t)^2 \, ,
\end{equation}

\noindent \eqref{eq:second_2} can be rewritten without the cross product,
\begin{equation} \label{eq:second_4}
    |\Be| \approx |\Bt| - \Ba \cdot \hat{B}_t 
    + \frac{|\Ba|^2}{2 |\Bt|}
    - \frac{(\Ba \cdot \hat{B}_t)^2}{2 |\Bt|} \, ,
\end{equation}

\noindent although this requires $|\Ba|$, which is not directly measurable. Note that ignoring the $\frac{(\Ba \cdot \hat{B}_t)^2}{2 |\Bt|}$ term typically results in less than 0.05\% error. Given the complexity of fitting quadratic terms for a marginal accuracy enhancement, it is likely best to focus development efforts on other sources of inaccuracies.

\section{Vector Aircraft Field Estimate}

\noindent Although the Tolles-Lawson approach nominally produces a scalar magnetometer correction via \eqref{eq:perm}-\eqref{eq:tl_2}, it is possible to estimate the vector components of the aircraft field directly: 

\begin{align*}
    \Ba &= (\beta_1 + |\Bt|
    \begin{bmatrix}
        \beta_4 & \beta_5/2 & \beta_6/2
    \end{bmatrix}
    \hat{B}_t + |\dot{\B}_t|
    \begin{bmatrix}
        \beta_{10} & \beta_{11} & \beta_{12}
    \end{bmatrix}
    \hat{\dot{B}}_t) \: \hat{i} \: + \\
    &\quad \: (\beta_2 + |\Bt|
    \begin{bmatrix}
        \beta_5/2 & \beta_7 & \beta_8/2
    \end{bmatrix}
    \hat{B}_t + |\dot{\B}_t|
    \begin{bmatrix}
        \beta_{13} & \beta_{14} & \beta_{15}
    \end{bmatrix}
    \hat{\dot{B}}_t) \: \hat{j} \: + \\
    &\quad \: (\beta_3 + |\Bt|
    \begin{bmatrix}
        \beta_6/2 & \beta_8/2 & \beta_9
    \end{bmatrix}
    \hat{B}_t + |\dot{\B}_t|
    \begin{bmatrix}
        \beta_{16} & \beta_{17} & \beta_{18}
    \end{bmatrix}
    \hat{\dot{B}}_t) \: \hat{k} \, .
\end{align*}

\noindent This vector estimate could ostensibly be used directly with vector magnetometers of sufficient quality.

\bigskip
\bigskip
\bigskip

\section*{Acknowledgments}

Research was sponsored by the United States Air Force Research Laboratory and the United States Air Force Artificial Intelligence Accelerator and was accomplished under Cooperative Agreement Number FA8750-19-2-1000. The views and conclusions contained in this document are those of the authors and should not be interpreted as representing the official policies, either expressed or implied, of the United States Air Force or the U.S. Government. The U.S. Government is authorized to reproduce and distribute reprints for Government purposes notwithstanding any copyright notation herein.

\bigskip
\bigskip
\bigskip

\bibliographystyle{IEEEtran}
\bibliography{main.bib}

\begin{thebibliography}{10}
\providecommand{\url}[1]{#1}
\csname url@samestyle\endcsname
\providecommand{\newblock}{\relax}
\providecommand{\bibinfo}[2]{#2}
\providecommand{\BIBentrySTDinterwordspacing}{\spaceskip=0pt\relax}
\providecommand{\BIBentryALTinterwordstretchfactor}{4}
\providecommand{\BIBentryALTinterwordspacing}{\spaceskip=\fontdimen2\font plus
\BIBentryALTinterwordstretchfactor\fontdimen3\font minus
  \fontdimen4\font\relax}
\providecommand{\BIBforeignlanguage}[2]{{%
\expandafter\ifx\csname l@#1\endcsname\relax
\typeout{** WARNING: IEEEtran.bst: No hyphenation pattern has been}%
\typeout{** loaded for the language `#1'. Using the pattern for}%
\typeout{** the default language instead.}%
\else
\language=\csname l@#1\endcsname
\fi
#2}}
\providecommand{\BIBdecl}{\relax}
\BIBdecl

\bibitem{Tolles1950}
W.~E. Tolles and J.~D. Lawson, ``{Magnetic compensation of MAD equipped
  aircraft},'' \emph{Report 201-1}, 1950.

\bibitem{Tolles1954}
\BIBentryALTinterwordspacing
W.~E. Tolles, ``{Compensation of Aircraft Magnetic Fields},'' US Patent
  2,692,970, pp. 1--8, 1954. [Online]. Available:
  \url{https://patents.google.com/patent/US2692970A}
\BIBentrySTDinterwordspacing

\bibitem{Tolles1955}
\BIBentryALTinterwordspacing
------, ``{Magnetic Field Compensation System},'' US Patent 2,706,801, pp.
  1--5, 1955. [Online]. Available:
  \url{https://patents.google.com/patent/US2706801A}
\BIBentrySTDinterwordspacing

\bibitem{Leliak1961}
\BIBentryALTinterwordspacing
P.~Leliak, ``{Identification and Evaluation of Magnetic-Field Sources of
  Magnetic Airborne Detector Equipped Aircraft},'' \emph{IRE Transactions on
  Aeronautical and Navigational Electronics}, vol. ANE-8, no.~3, pp. 95--105,
  1961. [Online]. Available: \url{https://doi.org/10.1109/TANE3.1961.4201799}
\BIBentrySTDinterwordspacing

\bibitem{Leach1979}
B.~W. Leach, ``{Automatic aeromagnetic compensation},'' \emph{Technical Report
  LTR-FR-69}, 1979.

\bibitem{Gu2013}
\BIBentryALTinterwordspacing
B.~Gu, Q.~Li, and H.~Liu, ``{Aeromagnetic Compensation Based on Truncated
  Singular Value Decomposition With an Improved Parameter-choice Algorithm},''
  in \emph{6th International Congress on Image and Signal Processing}.\hskip
  1em plus 0.5em minus 0.4em\relax IEEE, 2013, pp. 1545--1551. [Online].
  Available: \url{https://doi.org/10.1109/CISP.2013.6743921}
\BIBentrySTDinterwordspacing

\bibitem{Webb2014}
\BIBentryALTinterwordspacing
A.~C. White and M.~L. Webb, ``Environmental noise reduction for magnetometry,''
  US Patent 8\,786\,277, 2014. [Online]. Available:
  \url{https://patents.google.com/patent/US8786277}
\BIBentrySTDinterwordspacing

\bibitem{Wu2018a}
\BIBentryALTinterwordspacing
P.~Wu, Q.~Zhang, L.~Chen, W.~Zhu, and G.~Fang, ``{Aeromagnetic compensation
  algorithm based on principal component analysis},'' \emph{Journal of
  Sensors}, pp. 1--7, 2018. [Online]. Available:
  \url{https://doi.org/10.1155/2018/5798287}
\BIBentrySTDinterwordspacing

\bibitem{Bickel1979}
\BIBentryALTinterwordspacing
S.~H. Bickel, ``{Small Signal Compensation of Magnetic Fields Resulting from
  Aircraft Maneuvers},'' \emph{IEEE Transactions on Aerospace and Electronic
  Systems}, vol. AES-15, no.~4, pp. 518--525, 1979. [Online]. Available:
  \url{https://doi.org/10.1109/TAES.1979.308736}
\BIBentrySTDinterwordspacing

\bibitem{Larsson2007}
J.~Larsson, ``Electromagnetics from a quasistatic perspective,'' \emph{American
  Journal of Physics}, vol.~75, no.~3, pp. 230--239, mar 2007.

\bibitem{Griffiths1999}
D.~J. Griffiths, \emph{Introduction to Electrodynamics.}, 3rd~ed.\hskip 1em
  plus 0.5em minus 0.4em\relax Prentice Hall, 1999.

\bibitem{Herczynski2013}
A.~Herczy{\'{n}}ski, ``Bound charges and currents,'' \emph{American Journal of
  Physics}, vol.~81, no.~3, pp. 202--205, mar 2013.

\bibitem{Canciani2021}
\BIBentryALTinterwordspacing
A.~J. Canciani, ``{Magnetic Navigation on an F-16 Aircraft using Online
  Calibration},'' \emph{IEEE Transactions on Aerospace and Electronic Systems},
  pp. 1--15, 2021. [Online]. Available:
  \url{https://doi.org/10.1109/TAES.2021.3101567}
\BIBentrySTDinterwordspacing

\bibitem{Reeves2005}
C.~Reeves, \emph{{Aeromagnetic Surveys: Principles, Practice {\&}
  Interpretation}}.\hskip 1em plus 0.5em minus 0.4em\relax Geosoft
  Incorporated, 2005.

\bibitem{Canciani2016a}
\BIBentryALTinterwordspacing
A.~J. Canciani, ``{Absolute Positioning Using the Earth's Magnetic Anomaly
  Field},'' Doctoral dissertation, Air Force Institute of Technology, 2016.
  [Online]. Available: \url{https://scholar.afit.edu/etd/251/}
\BIBentrySTDinterwordspacing

\bibitem{Han2017}
\BIBentryALTinterwordspacing
Q.~Han, Z.~Dou, X.~Tong, X.~Peng, and H.~Guo, ``{A Modified Tolles–Lawson
  Model Robust to the Errors of the Three-Axis Strapdown Magnetometer},''
  \emph{IEEE Geoscience and Remote Sensing Letters}, vol.~14, no.~3, pp.
  334--338, 2017. [Online]. Available:
  \url{https://doi.org/10.1109/LGRS.2016.2640188}
\BIBentrySTDinterwordspacing

\bibitem{Noriega2011}
\BIBentryALTinterwordspacing
G.~Noriega, ``{Performance measures in aeromagnetic compensation},'' \emph{The
  Leading Edge}, vol.~30, no.~10, pp. 1122--1127, oct 2011. [Online].
  Available: \url{https://doi.org/10.1190/1.3657070}
\BIBentrySTDinterwordspacing

\bibitem{Drela2013}
M.~Drela, \emph{{Flight Vehicle Aerodynamics}}.\hskip 1em plus 0.5em minus
  0.4em\relax MIT Press, 2013.

\bibitem{Bickel1979a}
\BIBentryALTinterwordspacing
S.~H. Bickel, ``{Error Analysis of an Algorithm for Magnetic Compensation of
  Aircraft},'' \emph{IEEE Transactions on Aerospace and Electronic Systems},
  vol. AES-15, no.~5, pp. 620--626, 1979. [Online]. Available:
  \url{https://doi.org/10.1109/TAES.1979.308850}
\BIBentrySTDinterwordspacing

\bibitem{Nelson2020}
J.~B. Nelson, ``{Analysis of the Sander Geophysics Flt1002 and Flt1003 data},''
  Aeromagnetic Solutions Incorporated, Gloucester, ON, Canada, Tech. Rep.,
  2020.

\bibitem{Gnadt2022b}
\BIBentryALTinterwordspacing
A.~R. Gnadt, ``{Advanced Aeromagnetic Compensation Models for Airborne Magnetic
  Anomaly Navigation},'' Doctoral dissertation, Massachusetts Institute of
  Technology, 2022. [Online]. Available:
  \url{https://dspace.mit.edu/handle/1721.1/145137}
\BIBentrySTDinterwordspacing

\end{thebibliography}

\end{document}